\documentclass[10pt,twocolumn,letterpaper]{article}

\usepackage[pagenumbers]{cvpr} 

\usepackage{graphicx}
\usepackage{amsmath}
\usepackage{amssymb}
\usepackage{booktabs}
\usepackage{tikz}
\usepackage{multirow}
\usepackage{booktabs}
\usepackage{afterpage}
\usepackage{float}

\definecolor{cvprblue}{rgb}{0.21,0.49,0.74}
\usepackage[pagebackref,breaklinks,colorlinks,allcolors=cvprblue]{hyperref}

\def\paperID{12030} 
\def\confName{CVPR}
\def\confYear{2025}

\title{SIMPLE: Simultaneous Multi-Plane Self-Supervised Learning for Isotropic MRI Restoration from Anisotropic Data}

\author{Rotem Benisty, Yevgenia Shteynman, Moshe Porat, Moti Freiman\\ Technion – Israel Institute of Technology.\\ Haifa, Israel.\\
{\tt\small be.rotem@campus.technion.ac.il}
\and
Anat Ilivitzki\\
Rambam Health Care Campus. \\
Haifa, Israel.
}

\begin{document}
\maketitle
\begin{abstract}
Magnetic resonance imaging (MRI) is crucial in diagnosing various abdominal conditions and anomalies. Traditional MRI scans often yield anisotropic data due to technical constraints, resulting in varying resolutions across spatial dimensions, which limits diagnostic accuracy and volumetric analysis. Super-resolution (SR) techniques aim to address these limitations by reconstructing isotropic high-resolution images from anisotropic data. However, current SR methods often depend on indirect mappings and scarce 3D isotropic data for training, primarily focusing on two-dimensional enhancements rather than achieving genuine three-dimensional isotropy.
We introduce ``SIMPLE,'' a Simultaneous Multi-Plane Self-Supervised Learning approach for isotropic MRI restoration from anisotropic data. Our method leverages existing anisotropic clinical data acquired in different planes, bypassing the need for simulated downsampling processes. By considering the inherent three-dimensional nature of MRI data, SIMPLE ensures realistic isotropic data generation rather than solely improving through-plane slices. This approach's flexibility allows it to be extended to multiple contrast types and acquisition methods commonly used in clinical settings.
Our experiments on two distinct datasets (brain and abdomen) show that SIMPLE outperforms state-of-the-art methods both quantitatively using the Kernel Inception Distance (KID), semi-quantitatively through radiologist evaluations, and qualitatively through Fourier domain analysis. The generated isotropic volume facilitates more accurate volumetric analysis and 3D reconstructions, promising significant improvements in clinical diagnostic capabilities. Our code base will be available upon acceptance.
\end{abstract}    
\section{Introduction}
\label{sec:intro}
\begin{figure*}[t!]
  \centering
  \includegraphics[width=0.9\linewidth]{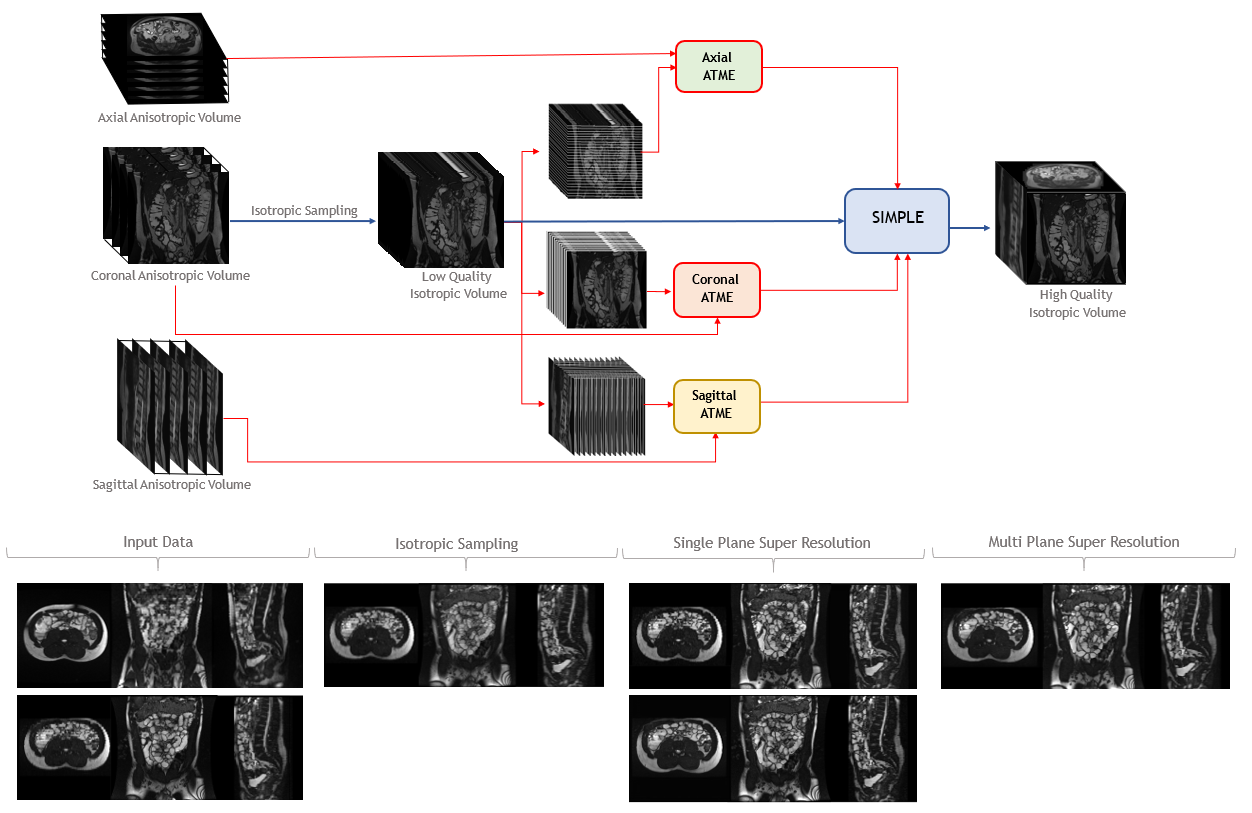}
  
   \caption{Our approach integrates ATME as a single-plane deep-learning-based super-resolution model \cite{solano2023look} and employs a simultaneous multi-plane super-resolution model to achieve isotropic resolution. The red arrows represent components utilized exclusively during the training phase in the model architecture. In contrast, the blue arrow denotes the primary path of the model, used during both the training and inference phases.}
   \label{fig:model_overview}
\end{figure*}

\label{sec:intro}



Magnetic resonance imaging (MRI) is essential in medical diagnostics, offering high-resolution, high-contrast images of soft tissues. In abdominal imaging, MRI is crucial for diagnosing and monitoring conditions like liver cirrhosis, pancreatic tumors, and renal anomalies~\cite{donato2017liver,gourtsoyiannis2011clinical,nowak2021detection,harrington2021mri,francis2023magnetic}. However, traditional MRI often yields anisotropic data with differing resolutions across spatial dimensions, limiting diagnostic accuracy and precise volumetric analysis~\cite{dev2023effect,ristow2009isotropic}.

This anisotropy stems from technical constraints in MRI, where achieving high spatial resolution requires lengthy acquisition times due to extensive ``k-space'' sampling~\cite{bankman2008handbook}. In clinical practice, time is limited by patient comfort, potential motion artifacts, and high throughput demands. In abdominal imaging, these challenges are intensified by motion from respiration and peristalsis, leading to compromises with higher in-plane (x, y) resolution and lower through-plane (z) resolution.



To reduce anisotropy and improve diagnostic accuracy, clinicians often acquire at least two high-resolution volumes in different planes, typically axial and coronal~\cite{gore2021textbook}. This multi-plane approach provides a more complete view, but partial volume effects and lower quality in the slice-selection direction still limit volumetric analysis and 3D reconstructions. Additionally, acquiring two planes extends scan time.

Super-resolution (SR) techniques offer a solution by reconstructing isotropic high-resolution images from anisotropic data~\cite{jia2017new}. Using advanced algorithms, often neural networks, SR methods learn to map low-resolution (large inter-slice spacing) images to high-resolution (small inter-slice spacing) images, enhancing anatomical detail, diagnostic accuracy, surgical planning, and quantitative assessments.

\begin{figure*}[t!]
  \centering
  \includegraphics[width=0.9\linewidth]{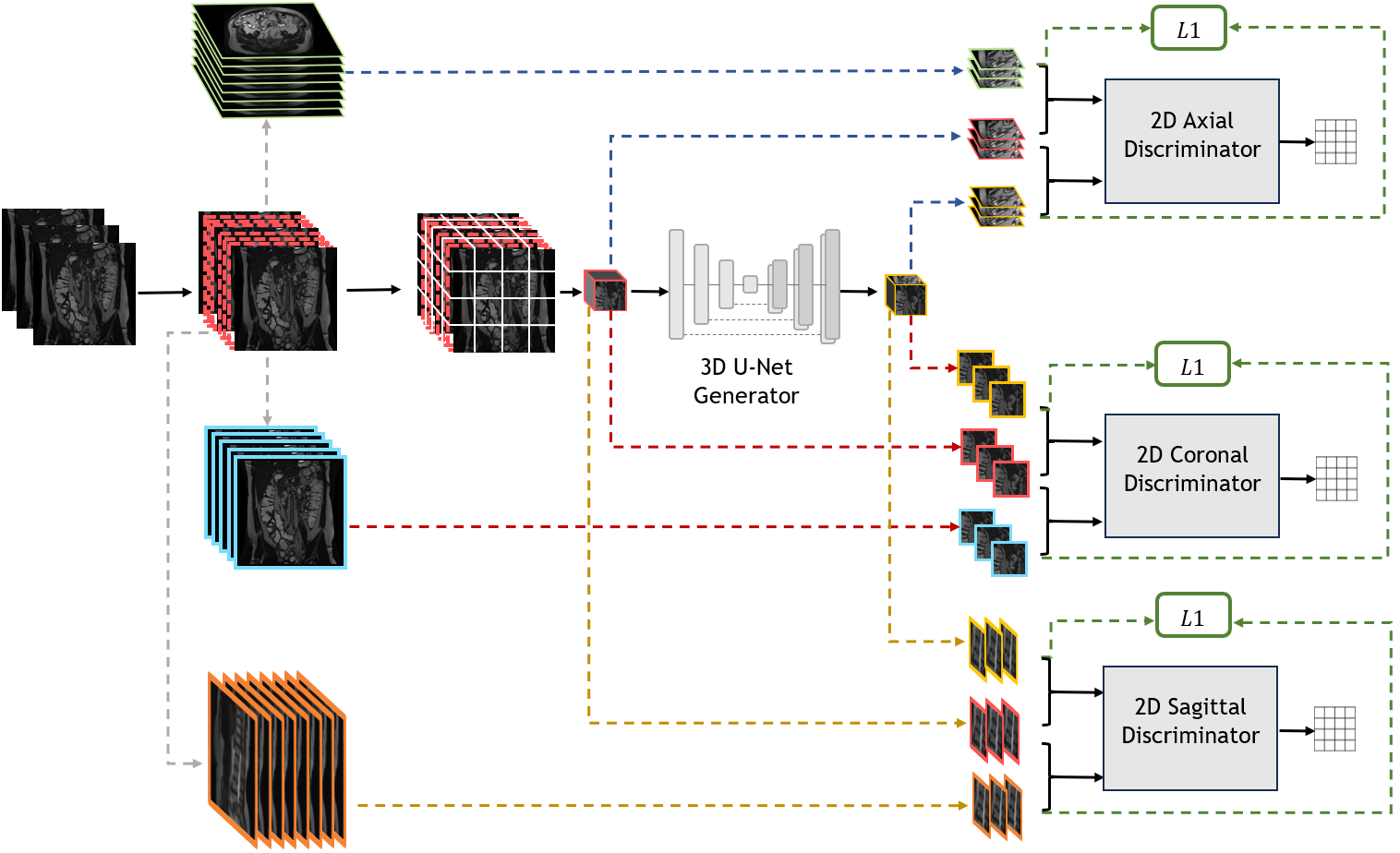}
  \caption{Model Architecture: The dashed red volume denotes the linearly interpolated volume. The blue-, green-, and orange-framed images represent the ATME-generated high-quality coronal, axial, and sagittal slices, respectively. Red, blue, and yellow dashed arrows indicate sampling along the coronal, axial, and sagittal planes, respectively. Gray dashed arrows show the ATME flow and green dashed arrows highlight the connections to consistency loss.}
   \label{fig:model_arch}
\end{figure*}



The abdomen presents unique MRI challenges due to motion artifacts from respiration and peristalsis, as well as the need for high contrast between structures. This makes acquiring isotropic data for training impractical, especially in clinical abdominal imaging, highlighting the need for SR methods that can handle anisotropic data.

To bypass the requirement for high-resolution (HR) training data, recent studies have developed self-supervised SR methods, categorized into resampling-based and synthesis-based techniques. Resampling-based methods simulate even lower resolution images from low-resolution (LR) data, map these back to the original LR images, and then use this mapping to predict HR images~\cite{jog2016self}. In contrast, synthesis-based methods generate HR images from LR data and train the SR model on these synthesized images~\cite{zhao2020smore,remedios2023self,lu2021two}.



Current self-supervised SR methods face limitations due to small training datasets and indirect mappings. Typically, these methods rely on a limited number of LR cases, learning SR mappings from either lower-resolution to low-resolution data or synthesized LR to synthesized HR, rather than real LR-HR pairs, which can impact model performance. Moreover, they generally focus on interpolating missing slices without fully considering the 3D structure of the data.

In this work, we introduce ``SIMPLE,'' a Simultaneous Multi-Plane Self-Supervised Learning approach for Isotropic MRI Restoration from Anisotropic Data. SIMPLE leverages anisotropic clinical scans from multiple planes and addresses the 3D nature of the data, beyond just improving through-plane slice quality. This flexible approach, without requiring specialized data, supports various contrast types and acquisition methods common in clinical practice. Figure~\ref{fig:model_overview} illustrates our model architecture.

\begin{figure*}[t!]
  \centering
  \begin{tikzpicture}
        \node (img) {\includegraphics[width=0.92\linewidth]{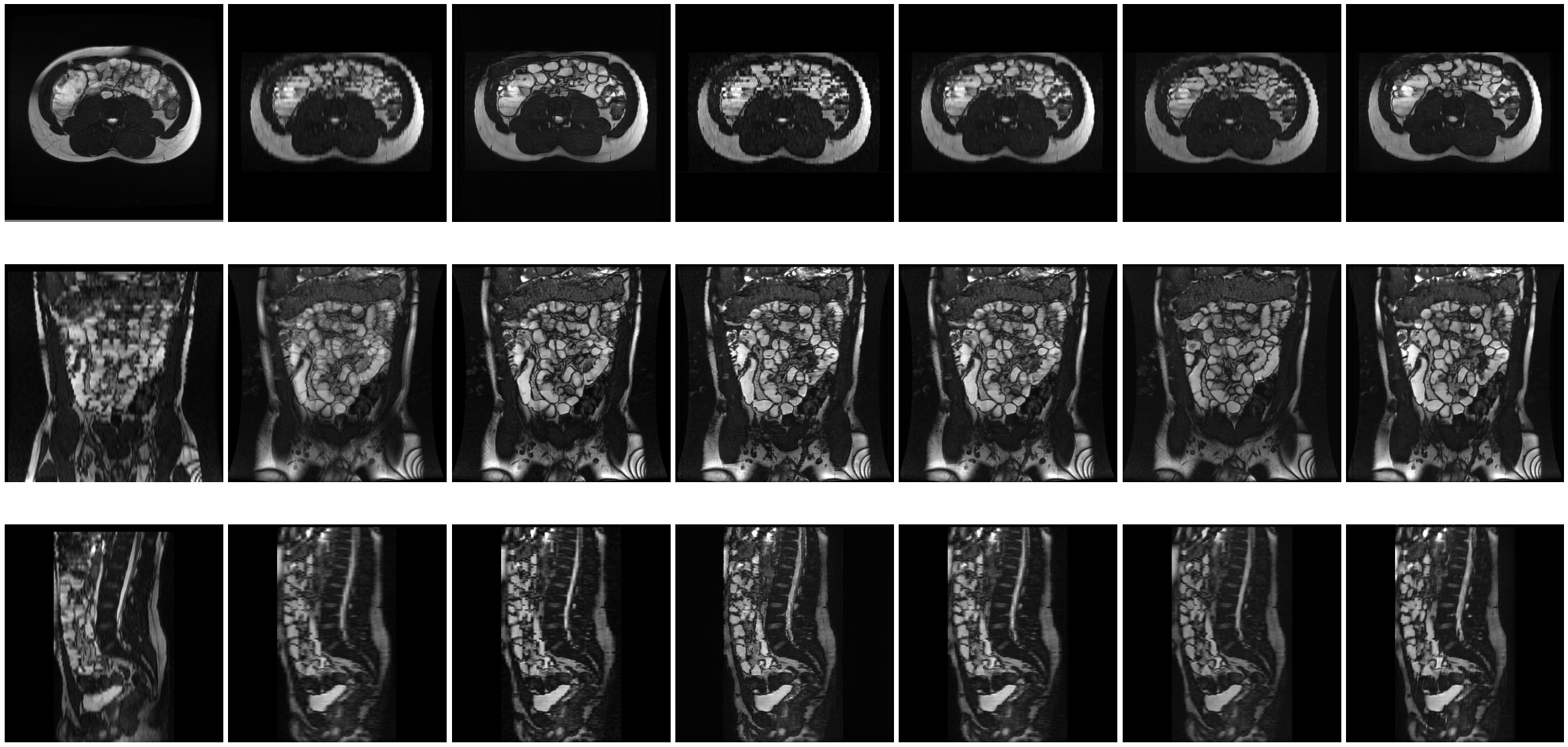}};

        \node[above] at ([xshift=-194,yshift=1pt]img.north) {\footnotesize Axial Volume};
        \node[above] at ([xshift=-129pt,yshift=1pt]img.north) {\footnotesize Coronal Volume};
        \node[above] at ([xshift=-65,yshift=1pt]img.north) {\footnotesize Axial ATME};
        \node[above] at ([xshift=1,yshift=1pt]img.north) {\footnotesize Coronal ATME};
        \node[above] at ([xshift=63,yshift=0pt]img.north) {\footnotesize Avg. ATME};
        \node[above] at ([xshift=129,yshift=1pt]img.north) {\footnotesize SMORE4};
        \node[above] at ([xshift=194,yshift=1pt]img.north) {\footnotesize SIMPLE};
        \node[left] at ([yshift=77pt]img.west) {\footnotesize \shortstack{Axial \\ View}};
        \node[left] at (img.west) {\footnotesize \shortstack{Coronal \\ View}};
        \node[left] at ([yshift=-80pt]img.west) {\footnotesize \shortstack{Sagittal \\ View}};
        
   \end{tikzpicture}
   \caption{Multi-plane view of slices sampled from the isotropic MRI volume generated by SIMPLE in comparison to five competitive methods (from left to right): linear interpolation on anisotropic axial volume, linear interpolation on anisotropic coronal volume, ATME on anisotropic axial volume, ATME on anisotropic coronal volume, averaged ATME on coronal and axial planes of anisotropic coronal volume, SMORE4 on anisotropic coronal volume, and SIMPLE on anisotropic coronal volume. SIMPLE produced higher-quality slices simultaneously on all planes compared to the other methods, which produced high-quality slices along the super-resolution plane.}
   \label{fig:multi_view}
\end{figure*}

\section{Previous Work}
\label{sec:previous_work}
The use of deep learning for through-plane super-resolution in MRI, which learns the non-linear mapping between low- and high-resolution images, has advanced significantly. Early supervised methods required high-resolution training data~\cite{masutani2020deep, peng2021vsr, wang2024interslicesuperresolutionmagneticresonance}, but these are often impractical in clinical settings, particularly for abdominal imaging. To overcome this, unsupervised methods have been developed. For example, Jog et al.~\cite{jog2016self} introduced a technique that downscales the LR plane further to create an LR$^2$ dataset, forming LR-LR$^2$ pairs to learn the mapping from LR$^2$ to LR, which is then applied to improve LR resolution.


Slice-to-volume (SVR)~\cite{uus2020deformable} and patch-to-volume (PVR)~\cite{alansary2017pvr} methods integrate axial, coronal, and sagittal images into an isotropic 3D volume. Yucheng et al.~\cite{liu20213d} proposed a framework for prostate MRI that applies PVR/SVR reconstruction to generate an initial isotropic volume, then re-slices it into 2D slices for SR training across all planes. The three SR volumes are then averaged, yielding a final 3D SR volume with enhanced isotropic resolution.


Zhao et al.~\cite{zhao2020smore} used the HR plane in anisotropic volumes to create HR-LR pairs by downsampling for their ``SMORE'' model, which is then applied to enhance the LR plane. Remedios et al.~\cite{remedios2023self} improved this by learning the point-spread-function (PSF) in the LR plane for more accurate downsampling simulation, creating HR-LR pairs and averaging restored volumes from two LR planes to produce a single HR isotropic volume. Similarly, Lu et al.~\cite{lu2021two} introduced TSCNet, a two-stage self-supervised learning method that pretrains on synthesized LR-HR pairs in sagittal and coronal planes, followed by cyclic interpolation using triplet axial slices for refinement.


Zhang et al.~\cite{zhang2023self} constructed HR-LR image pairs for training and adapted an implicit neural representation (INR) network for a 2D arbitrary-scale super-resolution (SR) model. Fang et al.~\cite{fang2024cycleinr} later introduced ``CycleINR,'' an enhanced 3D volumetric SR model that leverages the continuity of implicit functions for arbitrary up-sampling rates without separate training. They further improved performance with local attention for grid sampling and reduced over-smoothing through cycle-consistent loss.


Sander et al.~\cite{sander2022autoencoding} proposed an unsupervised deep learning approach for semantic interpolation, synthesizing intermediate slices from low-resolution inputs. By leveraging the latent space of autoencoders, they achieved smooth through-plane interpolation, generating new slices by convexly combining latent encodings of adjacent slices and decoding the result into an intermediate slice.


However, these methods assume a high-quality approximation of the downsampling process and focus on generating high-resolution images in the through-plane, neglecting the three-dimensional isotropic nature of the data. In contrast, our approach uses existing multi-plane anisotropic clinical data, bypassing simulated downsampling and enhancing quality across all planes to capture the full 3D structure.


\section{Method}
\label{sec:method}
To address the limitations of previous works and enable simultaneous multi-plane processing, we introduce SIMPLE—a simultaneous multi-plane self-supervised learning model for isotropic super-resolution MRI volume reconstruction from anisotropic MRI data acquired using 2D MRI techniques. Our model effectively manages slice spacing and motion, enhancing the resolution and quality of the 3D MRI volume. This improvement ensures that sampling along any plane (axial, coronal, or sagittal) results in high-quality 2D slices in all planes.
SIMPLE takes as input an anisotropic volume, and produces a low-quality isotropic volume using a linear interpolation. Then, the model restores a high-quality isotropic volume from the low-quality one. Equation~\ref{eq:model} provides a mathematical framework for the proposed model. $V_{\text{An-Iso}}$ denotes the anisotropic volume, $V'_{\text{Iso}}$ represents the linearly interpolated low-quality isotropic volume, $\hat{V}_{\text{Iso}}$ signifies the output high-quality isotropic volume, $GM$ denotes the generative model and $L$ the linear interpolation operator. This formulation distinguishes our approach by addressing all three constraints, unlike previous methods which typically focus on resolving only one or two.

\begin{equation}
\begin{aligned}
&\hat{V}_{\text{Iso}} = GM(V'_{\text{Iso}}) = GM(L(V_{\text{An-Iso}})) \\
s.t: \ \ \ \ \ \ \ &\text{I. } S_{\text{HR}_{\text{Cor}}} = \hat{V}_{\text{Iso}}[i, :, :] = HR(V'_{\text{Iso}}[i, :, :]) \\
&\text{II. } S_{\text{HR}_{\text{Ax}}} = \hat{V}_{\text{Iso}}[:, i, :] = HR(V'_{\text{Iso}}[:, i, :]) \\
&\text{III. } S_{\text{HR}_{\text{Sag}}} = \hat{V}_{\text{Iso}}[:, :, i] = HR(V'_{\text{Iso}}[:, :, i])
\end{aligned}
\label{eq:model}
\end{equation}

We implemented our approach by combining a 3D generator based on the U-Net architecture~\cite{cciccek20163d} with 2D conditional patch discriminators~\cite{isola2017image}, specifically designed for the coronal, axial and sagittal planes. Figure~\ref{fig:model_arch} depicts the overall architecture.

\subsection{Model Architecture}
\subsubsection{Pre-Processing}
\label{sec:preprocessing}
As a pre-processing step, we isotropically sample the anisotropic input volume by performing linear interpolation. Due to the significant spacing between the MRI slices, linear interpolation struggles to accurately predict the voxels between the slices, resulting in a low-quality 3D volume. This is reflected in the poor quality of slices in all main 2D planes. To manage the high computational demands of processing the uniform isotropic volume, we extracted 3D patches of size 64$\times$64$\times$64 with an overlap in each dimension.
\subsubsection{Networks}
We employ a 3D domain generator based on the 3D U-Net architecture~\cite{cciccek20163d} that takes a 64$\times$64$\times$64 single-channel, low-quality 3D patch as input and outputs a high-quality patch of the same dimensions. This convolutional neural network has a contracting path for capturing context and an expanding path for precise localization, making it effective for tasks where feature location is critical. The encoder-decoder architecture includes skip connections: the encoder has multiple convolutional layers (kernel size 4, stride 2) with LeakyReLU activation and instance normalization, while the decoder uses transposed convolution layers (kernel size 4, stride 2) with ReLU activation, instance normalization, and dropout layers.

We use two or three discriminators, one for each plane. The number of discriminators depends on data availability: three discriminators are used when the dataset includes anisotropic images from all three planes, while only two discriminators are used if the data includes images from only two planes. Unlike the 3D generator, these discriminators operate in the 2D domain. We extract 3D input and output patches along all planes from the generator, creating 64 2D patches of size 64$\times$64 from each 3D patch. Inspired by the pix2pix model~\cite{isola2017image}, each plane has a conditional patch discriminator that takes the input image and either the generated or real image as a single 2-channel input, ensuring the generated image remains realistic and consistent.

The patch discriminator focuses on local features and textures by dividing the image into patches and evaluating their authenticity independently. Real and synthetic samples are input as 2D patches with two channels: the first channel is a patch from the low-quality isotropic 3D input to the generator, and the second channel differs between real and synthetic samples. For synthetic samples, it is a 2D patch from the generator's 3D output, while for real samples, it is a 2D patch from a high-resolution isotropic MRI volume in a specific plane.

All discriminators use 2D convolutional layers (kernel size 4, stride 2), followed by instance normalization and LeakyReLU activation. Since the volume's original orientation is in a specific plane, the generator's task is simpler in this plane compared to others. To increase complexity in the original plane, we use a simpler architecture for the corresponding discriminator, with fewer layers and filters.

\subsubsection{Loss Functions}
For the SIMPLE generator, we employ a loss function that equally integrates coronal, axial and sagittal components. Each component comprises two types of loss: adversarial loss $\mathcal{L}_{\text{G}_{\text{ADV}}}$, derived from the discriminator's output classification matrix, and consistency loss $\mathcal{L}_{L_1}$, calculated as the L1 distance. For the discriminators, we use only the adversarial loss. The adversarial loss is based on a least square GAN (LSGAN), which provides more stable training compared to vanilla GANs, and conditional patch discriminator. Equation~\ref{eq:cLSGAN_G} represents the LSGAN loss for the generator, where $x$ denotes the input image. 
\begin{equation}
\small
\mathcal{L}_{\text{G}_{\text{ADV}}} = \mathcal{L}_{\text{LSGAN}}(G,D) = \mathbb{E}_{x} \left[ \left( D(x, G(x)) - 1 \right)^2 \right]
\label{eq:cLSGAN_G}
\end{equation}
Equation~\ref{eq:cLSGAN_D} represents the LSGAN loss for the discriminator, where $y$ denotes the ground truth image corresponding to the input image $x$.
\begin{equation}
\begin{split}
\small
\mathcal{L}_{\text{D}_{\text{ADV}}} = \mathcal{L}_{\text{LSGAN}}(D) = &\frac{1}{2} \mathbb{E}_{x,y} \left[ \left( D(x, y) - 1 \right)^2 \right] + \\ 
                                 &\frac{1}{2} \mathbb{E}_{x} \left[ \left( D(x, G(x)) \right)^2 \right]
\label{eq:cLSGAN_D}
\end{split}
\end{equation}
By incorporating consistency loss, the generator aims not only to deceive the discriminator but also to closely approximate the ground truth output. We choose L1 distance because it encourages sparsity and sharp edges in the generated image, effectively reducing blurring. Equation~\ref{eq:L1} represents the L1 distance metric between the generated 2D patches and the super-resolution 2D patches, with $\lambda$ denoting the weight for this loss. We set $\lambda$ to 10.
\begin{equation}
\small
\begin{split}
\mathcal{L}_{L_1}(G) = \lambda \mathbb{E}_{x,y} \left[ \|y - G(x)\|_1 \right]
\label{eq:L1}
\end{split}
\end{equation}
Equation~\ref{eq:total_L} represents our total generator loss:
\begin{equation}
\small
\begin{split}
\mathcal{L}_{\text{G}} = &\ \alpha \left[ \mathcal{L}_{\text{G}_{\text{ADV}_{\text{cor}}}} + \lambda_{\text{cor}} \mathcal{L}_{L_{1_{\text{cor}}}} \right] 
+ \beta \left[ \mathcal{L}_{\text{G}_{\text{ADV}_{\text{ax}}}} + \lambda_{\text{ax}} \mathcal{L}_{L_{1_{\text{ax}}}} \right] 
\\
&\ \ \ \ \ \ \ \ \ \ \ + \gamma \left[ \mathcal{L}_{\text{G}_{\text{ADV}_{\text{sag}}}} + \lambda_{\text{sag}} \mathcal{L}_{L_{1_{\text{sag}}}} \right]
\label{eq:total_L}
\end{split}
\end{equation}
where $\alpha$, $\beta$, $\gamma$ denote the weights of the coronal, axial and sagittal components, respectively. We set $\alpha$, $\beta$ and $\gamma$ both to 0.5.

\subsection{Training}
Figure~\ref{fig:model_arch} illustrates the training process step-by-step using the dashed arrows. The key components of the training involve isotropic sampling, detailed in section~\ref{sec:preprocessing}, performing single-plane super-resolution on the coronal, axial and sagittal planes as an initial step to generate paired data for the conditional model, and ultimately, training the generator and discriminators.

\subsubsection{Single Plane Super Resolution}
Although our databases contain MRI sequences in two or three planes for each case, these sequences were scanned at different times. Consequently, the physiological motion organs (such as the bowel in the digestive tract) and patient movement during scanning can cause shifts in the area within the field of view (FOV) of these sequences. This leads to motion artifacts and results in incomplete and inconsistent coverage of the patient anatomy between these two sequences. As a result, we cannot rely on these sequences as ground truth for our model and must address the problem using an unsupervised method. However, to still provide real samples to our model, we perform a single image deep-learning-based super-resolution model on the interpolated isotropic low-resolution MRI volume for each plane separately. This approach yields isotropic volumes, each with high resolution in only one plane. For the single-image super-resolution model, we choose ATME, a recent approach for image-to-image translation tasks that integrates elements from GAN and diffusion models \cite{solano2023look}. We train two ATME models, each on a different plane, and evaluate them on slice samples from the interpolated isotropic volume in the coronal, axial and sagittal planes.   

\subsubsection{Training Setup}
The generator and the discriminators are trained together for 100 epochs with a batch size of 16. The learning rate is initially set to 0.0002 and updated using a step scheduler. The Adam optimizer is employed to update the model parameters, with beta1 set to 0.5 and beta2 set to 0.999.

\begin{figure}[]
  \centering
  \begin{tikzpicture}
        \node (img) {\includegraphics[width=0.9\linewidth]{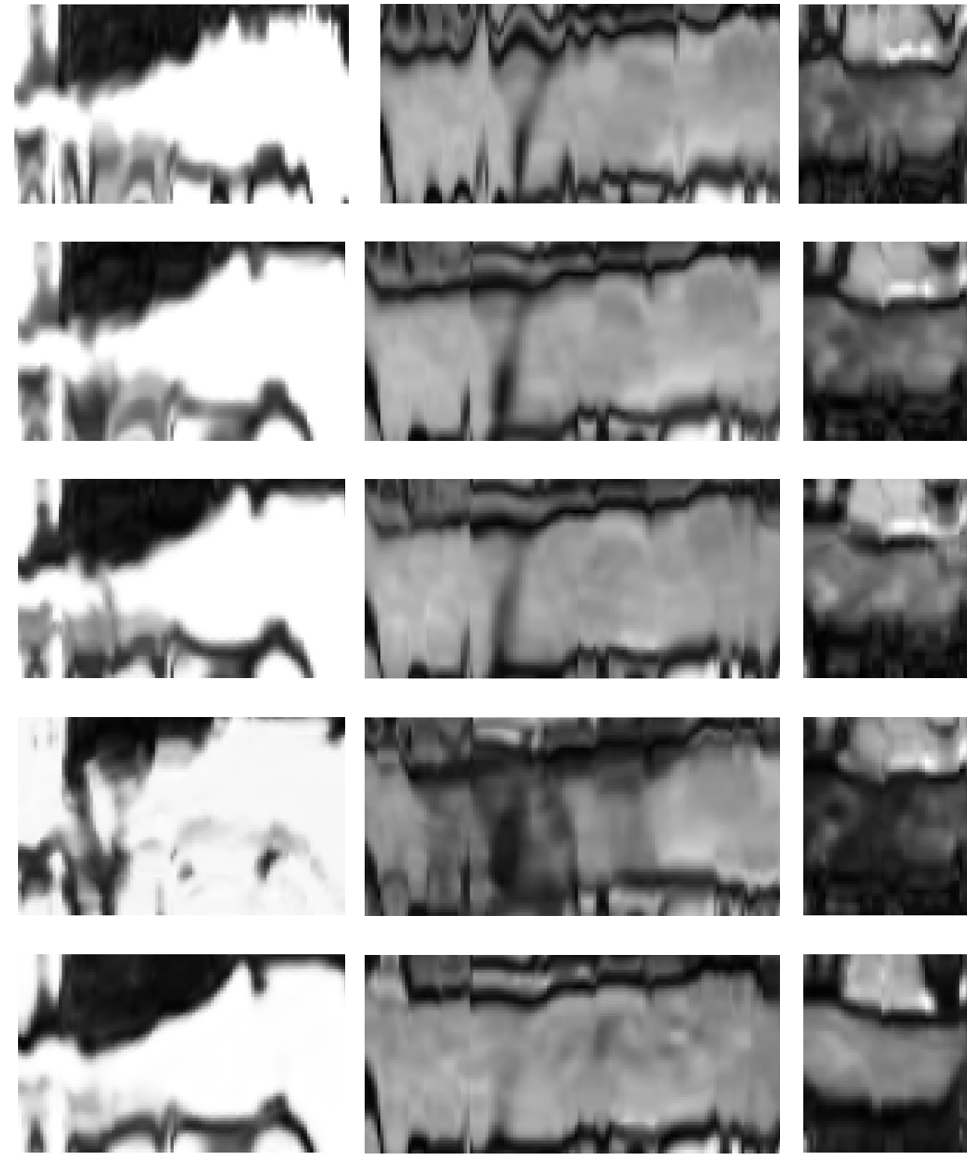}};

        \node[below] at ([xshift=-72]img.south) {\footnotesize case 1};
        \node[below] at ([xshift=20]img.south) {\footnotesize case 2};
        \node[below] at ([xshift=93]img.south) {\footnotesize case 3};

        \node[left] at ([yshift=100pt]img.west) {\footnotesize (a)};
        \node[left] at ([yshift=50pt]img.west) {\footnotesize (b)};
        \node[left] at ([yshift=-2pt]img.west) {\footnotesize (c)};
        \node[left] at ([yshift=-55pt]img.west) {\footnotesize (d)};
        \node[left] at ([yshift=-107pt]img.west) {\footnotesize (e)};
        
   \end{tikzpicture}
   \caption{Straight multi-planar reconstructions (MPR) of the terminal ileum in Crohn's disease patients for three cases based on coronal volumes: (a) anisotropic volume, (b) linearly interpolated isotropic volume, (c) averaged ATMEs isotropic volume, (d) SMORE4 isotropic volume, and (e) SIMPLE isotropic volume. The MPRs generated from the SIMPLE reconstructed volume were less prone to artifacts stemming from the anisotropic MRI acquisition.}
   \label{fig:StraightMPR}
\end{figure}

\section{Experiments}
\label{sec:experiments}
\subsection{Dataset}
We utilize two distinct datasets to evaluate our model across different anatomical organs, variations in spatial resolution and volume dimensions, as well as different levels of slice-to-slice motion artifacts.
\subsection*{Abdomen Dataset}
We obtained the appropriate Institutional Review Board (IRB) approval for retrospective data analysis. We collected 115 consecutive 2D MRI scans of Crohn's Disease (CD) patients from a local hospital. All the MRI scans were acquired with a GE Medical Systems 3T MRI scanner. For the study, we utilize the coronal and axial FIESTA sequences. In both Coronal and Axial cases, the slice spacing and slice thickness are both set at 5mm, ensuring there is no overlap between adjacent slices. The average pixel spacing for the coronal sequences ranged from 0.7 mm to 0.94 mm in the x and y directions (mean 0.78 mm), while for the axial sequences, it ranged from 0.66 mm to 0.94 mm (mean 0.76 mm). To standardize the isotropic volume to a uniform size of 512 in each dimension, we apply zero padding by adding empty slices along the edges of the coronal plane during the evaluation of our model.


\subsection*{Oasis-1 Dataset}
We use the publicly available OASIS-1 dataset, containing 416 cross-sectional MRI brain scans from adults across different age groups and cognitive statuses. Each subject's data includes 3-4 T1-weighted MRI scans in the sagittal plane, with 1.25 mm slice spacing and 1 mm pixel resolution. The dataset also provides an averaged, co-registered image resampled to an isotropic voxel size of 1 mm, serving as the 3D isotropic ground truth for our study. From this, we generate three anisotropic 2D MRI image sequences in the coronal, axial, and sagittal planes, with a slice spacing of 3 mm and a pixel resolution of 1 mm, by applying a Gaussian filter ($\sigma = 3$) and selecting every third slice. Zero padding is applied to standardize the size to 256 in each dimension.

\begin{figure}[t!]
  \centering
  \begin{tikzpicture}
        \node (img) {\includegraphics[width=1\linewidth]{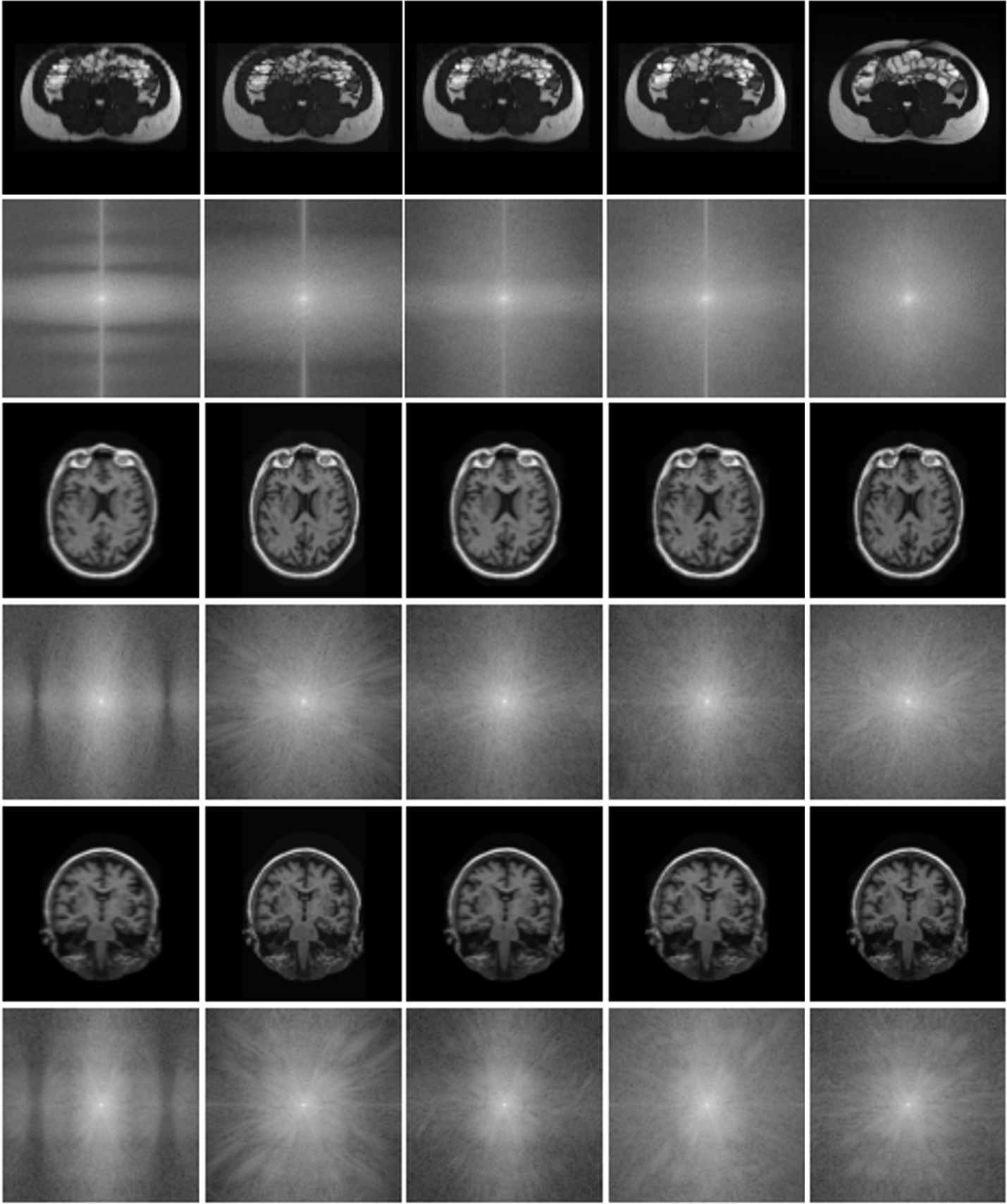}};

        \node[above] at ([xshift=-95]img.north) {\footnotesize Interpolation};
        \node[above] at ([xshift=-48,yshift=1pt]img.north) {\footnotesize SMORE4};
        \node[above] at ([xshift=0,yshift=0pt]img.north) {\footnotesize Avg. ATME};
        \node[above] at ([xshift=48,yshift=1pt]img.north) {\footnotesize SIMPLE};
        \node[above] at ([xshift=95,yshift=1pt]img.north) {\footnotesize GT Volume};

   \end{tikzpicture}
   \caption{Fourier representation of axial and coronal slices generated from an anisotropic axial and coronal volumes, accordingly. SIMPLE's Fourier representation shows a more consistent distribution of the Fourier coefficients across the two dimensions of the image, whereas other methods images exhibit sampling artifacts. Compared to the closest slice from the axial or coronal GT volume, SIMPLE produces a similar Fourier representation.}
   \label{fig:FFT}
\end{figure}


\subsection{Comparison Methods}
We compare SIMPLE against linear interpolation, SMORE4~\cite{remedios2023self}, and an approach similar to Yucheng et al.~\cite{liu20213d}, in which super-resolution 3D volumes are generated by ATME in each plane and then averaged. All methods were evaluated on identical anisotropic MRI volumes from the same plane using consistent performance metrics. For evaluation, we used the coronal plane for the abdomen dataset and the sagittal plane for the brain dataset.

\subsection{Qualitative Methods}
We assess SIMPLE’s qualitative performance through visual assessments, multi-planar reconstructions, Fourier domain analysis, and senior radiologists feedback.



\subsubsection*{Multi View Slices:} 
We examine sampled slices in the coronal, axial, and sagittal planes for SIMPLE and comparison methods, evaluated on anisotropic volumes. Unlike the brain dataset, the abdomen dataset lacks 3D isotropic volumes, so ground truth slices for comparison are not available.
\subsubsection*{Straight Multi Planar Reconstruction (MPR):}  For the abdomen dataset, we use an in-house application to create straight multi-planar reconstructions (MPRs)  
\cite{kanitsar2002cpr} of the small intestine, allowing assessment of voxel correlations in the 3D domain. We utilized a dataset of Terminal Ileum (TI) centerlines from Crohn's Disease (CD) patients, manually annotated by an attending radiologist on the same MRI database from our local hospital.
\subsubsection*{Fourier Domain:} 
To analyze the frequency components of the generated slices, we apply the Fourier transform to sampled slices from the isotropic volume in planes other than the evaluation plane. This evaluates the preservation of high-frequency details, such as textures and edges, assessing the similarity between ground-truth and generated MRI images.
\subsubsection*{Likert Scale:} 
A senior radiologist ranked slices from 10 cases in our abdomen dataset across the coronal, axial, and sagittal views. These slices were sampled from isotropic volumes generated by SIMPLE, SMORE4, and linear interpolation. The radiologist, blinded to the methods used, assigned rankings on a Likert scale from 1 to 5: 1 for non-diagnostic images, 2 for poor quality, 3 for partial visualization of anatomy with blurry borders, 4 for good visualization with mild blurring, and 5 for excellent visualization.

\subsection{Quantitative Methods}
We quantitatively compared SIMPLE to other methods using distribution-based metrics. For the abdomen dataset we used Kernel Inception Distance (KID) and Inception Score (IS) \cite{barratt2018note}, derived from the InceptionV3 network. Additionally, we evaluated the Fr\'echet Inception Distance (FID)~\cite{kastryulin2023image} using the VGG16 architecture \cite{simonyan2014very} as the backbone. 

In their thorough review, Yi et al.~\cite{yi2019generative} highlight that standard metrics like PSNR and SSIM do not effectively capture the visual quality in GAN-generated medical images, raising concerns over the limitations of traditional full-reference metrics. Consequently, we chose to assess the brain dataset using the KID metric, which does not rely exclusively on ground-truth comparisons. We averaged the KID scores across all 2D slices in each plane. Additionally, we assessed 3D volume quality by averaging KID scores across 2D slices per plane, then averaging across cases.


\section{Results}
\label{sec:results}
\subsection{Comparison of Ours to Baselines}
\subsubsection*{Multi View Slices:} Figures ~\ref{fig:multi_view} presents views of the three primary planes of isotropic MRI volumes generated by all methods. SIMPLE effectively enhances quality and resolution in both axial and coronal planes, unlike methods that favor a specific plane. Interpolation and ATME improve the plane quality corresponding to their original anisotropic acquisition. SIMPLE, SMORE4 and averaged ATMEs all improve resolution and quality across all planes, but SIMPLE outperforms in the planes perpendicular to the acquisition plane by better detecting and sharpening edges and improving contrast. In the acquisition plane, the differences between SIMPLE and SMORE4 are subtle, with some organ locations differing. For the abdomen dataset, SMORE4 is sensitive to organ shadows and emphasizes noise. While SIMPLE has no constraints in the sagittal plane for the abdomen dataset, it improves resolution. For the brain dataset, differences between SIMPLE to SMORE4, and averaged ATMEs are less pronounced than in the abdomen dataset, due to smaller slice spacing and lower likelihood of organ motion.






\begin{table}[t]
\small
\resizebox{\columnwidth}{!}{
    \begin{tabular}{cc|c|c}
                                       & \textbf{Interpolation} & \textbf{SMORE4}   & \textbf{SIMPLE}                        \\ \cline{2-4} 
\multicolumn{1}{c|}{\textbf{Axial}}    & 1                      & 2.182 $\pm$ 1.168 & \multicolumn{1}{c|}{{\bf 3.727 $\pm$ 0.786}} \\ \hline
\multicolumn{1}{c|}{\textbf{Coronal}} & 1.231 $\pm$ 0.438 & 4.231 $\pm$ 0.438 & \multicolumn{1}{c|}{\bf 4.615 $\pm$ 0.65} \\ \hline
\multicolumn{1}{c|}{\textbf{Sagittal}} & 1                      & 1.5 $\pm$ 0.527   & \multicolumn{1}{c|}{{\bf 2.2 $\pm$ 0.789}}   \\ \cline{2-4} 
\end{tabular}
}
\caption{Average and standard deviation of the Likert scale ranking for all 3 main planes.}
\label{table:likert_scale}
\end{table}

\subsubsection*{Straight Multi Planar Reconstruction (MPR):} Figure~\ref{fig:StraightMPR} depicts three straight MPRs based on five volumes: the original anisotropic volume and isotropic volumes generated by all methods. SIMPLE's reconstruction appears smoother compared to both the anisotropic and interpolated isotropic volumes, which contain discontinuities due to the larger gap between slices and inadequate generation. In some cases, the straight MPRs of SMORE4 and averaged ATMEs appear blurrier than that of SIMPLE.

\subsubsection*{Fourier Domain:} Figure~\ref{fig:FFT} shows the Fourier domain of axial and coronal slices from isotropic volumes generated by all methods. For both datasets, SIMPLE's Fourier transform resembles that of natural images, with a balanced mix of high and low frequencies. In contrast, the interpolated axial abdominal slices exhibit horizontal stripes due to the anisotropic resolution in the axial plane. Similarly, SMORE4 and averaged ATMEs display a prominent horizontal stripe near the center. All methods show distinct vertical and horizontal lines around the center, likely caused by zero-padding applied to the isotropic volume. For the brain dataset, it is apparent that the coronal and axial slices of the competitive methods lack high-frequency components and exhibit a greater disparity from the Fourier domain of the original images.

\subsubsection*{Likert Scale:} Table~\ref{table:likert_scale} presents the average and standard deviation of Likert scale rankings for each main plane across all methods. As expected, linear interpolation received the lowest average rank. For the coronal plane, SMORE4 and SIMPLE achieved similar ranks, with SIMPLE having a slight advantage. In the axial plane, SIMPLE's average rank is significantly higher than SMORE4's. In the sagittal plane, both methods received low ranks, but SIMPLE's rank is higher.


\begin{table}[t]
\small
\resizebox{\columnwidth}{!}{
\begin{tabular}{ccc|cc|cc|cc}
 &
  \multicolumn{2}{c|}{\textbf{Interpolation}} &
  \multicolumn{2}{c|}{\textbf{SMORE4}} &
  \multicolumn{2}{c|}{\textbf{Avg. ATME}} &
  \multicolumn{2}{c}{\textbf{SIMPLE}} \\ \cline{2-9} 
\multicolumn{1}{c|}{} &
  \multicolumn{1}{c|}{\textit{Coronal}} &
  \textit{Axial} &
  \multicolumn{1}{c|}{\textit{Coronal}} &
  \textit{Axial} &
  \multicolumn{1}{c|}{\textit{Coronal}} &
  \textit{Axial} &
  \multicolumn{1}{c|}{\textit{Coronal}} &
  \multicolumn{1}{c|}{\textit{Axial}} \\ \cline{2-9} 
\multicolumn{1}{c|}{\multirow{2}{*}{\textbf{KID}}} &
  \multicolumn{1}{c|}{3.829} &
  48.185 &
  \multicolumn{1}{c|}{0.894} &
  40.554 &
  \multicolumn{1}{c|}{0.639} &
  37.213 &
  \multicolumn{1}{c|}{2.383} &
  \multicolumn{1}{c|}{{\bf 32.487}} \\ \cline{2-9} 
\multicolumn{1}{c|}{} &
  \multicolumn{2}{c|}{26.007} &
  \multicolumn{2}{c|}{20.724} &
  \multicolumn{2}{c|}{18.926} &
  \multicolumn{2}{c|}{{\bf 17.435}} \\ \hline
\multicolumn{1}{c|}{\multirow{2}{*}{\textbf{IS}}} &
  \multicolumn{1}{c|}{0.031} &
  0.023 &
  \multicolumn{1}{c|}{0.001} &
  0.033 &
  \multicolumn{1}{c|}{0.005} &
  0.022 &
  \multicolumn{1}{c|}{0.003} &
  \multicolumn{1}{c|}{{\bf 0.006}} \\ \cline{2-9} 
\multicolumn{1}{c|}{} &
  \multicolumn{2}{c|}{0.027} &
  \multicolumn{2}{c|}{0.017} &
  \multicolumn{2}{c|}{0.013} &
  \multicolumn{2}{c|}{{\bf 0.004}} \\ \hline
\multicolumn{1}{c|}{\multirow{2}{*}{\textbf{FID}}} &
  \multicolumn{1}{c|}{17.292} &
  28.964 &
  \multicolumn{1}{c|}{14.897} &
  23.938 &
  \multicolumn{1}{c|}{15.387} &
  22.969 &
  \multicolumn{1}{c|}{16.834} &
  \multicolumn{1}{c|}{{\bf 21.316}} \\ \cline{2-9} 
\multicolumn{1}{c|}{} &
  \multicolumn{2}{c|}{23.128} &
  \multicolumn{2}{c|}{19.417} &
  \multicolumn{2}{c|}{19.178} &
  \multicolumn{2}{c|}{{\bf 19.075}} \\ \cline{2-9} 
\end{tabular}
}
\caption{Distribution-based metrics for sampled axial and coronal slices from the generated isotropic volume in the abdomen dataset. It includes the scores for each plane individually and the average score for both planes. Lower scores indicate better performance.}
\label{table:res_abdomen}
\end{table}

\begin{table}[t]
\small
\resizebox{\columnwidth}{!}{
\begin{tabular}{cccl|ccl|ccl|ccl}
 &
  \multicolumn{3}{c|}{\textbf{Interpolation}} &
  \multicolumn{3}{c|}{\textbf{SMORE4}} &
  \multicolumn{3}{c|}{\textbf{Avg. ATME}} &
  \multicolumn{3}{c}{\textbf{SIMPLE}} \\ \cline{2-13} 
\multicolumn{1}{c|}{} &
  \multicolumn{1}{c|}{\textit{Cor}} &
  \multicolumn{1}{c|}{\textit{Ax}} &
  \textit{Sag} &
  \multicolumn{1}{c|}{\textit{Cor}} &
  \multicolumn{1}{c|}{\textit{Ax}} &
  \textit{Sag} &
  \multicolumn{1}{c|}{\textit{Cor}} &
  \multicolumn{1}{c|}{\textit{Ax}} &
  \textit{Sag} &
  \multicolumn{1}{c|}{\textit{Cor}} &
  \multicolumn{1}{c|}{\textit{Ax}} &
  \multicolumn{1}{l|}{\textit{Sag}} \\ \cline{2-13} 
\multicolumn{1}{c|}{\textbf{\begin{tabular}[c]{@{}c@{}}2D\\ KID\end{tabular}}} &
  \multicolumn{1}{c|}{0.89} &
  \multicolumn{1}{c|}{0.964} &
  \multicolumn{1}{c|}{0.386} &
  \multicolumn{1}{c|}{0.47} &
  \multicolumn{1}{c|}{0.495} &
  \multicolumn{1}{c|}{0.354} &
  \multicolumn{1}{c|}{0.418} &
  \multicolumn{1}{c|}{0.455} &
  \multicolumn{1}{c|}{0.241} &
  \multicolumn{1}{c|}{\bf 0.396} &
  \multicolumn{1}{c|}{\bf 0.422} &
  \multicolumn{1}{c|}{\bf 0.231} \\ \hline
\multicolumn{1}{c|}{\textbf{\begin{tabular}[c]{@{}c@{}}3D\\ KID\end{tabular}}} &
  \multicolumn{3}{c|}{30.425} &
  \multicolumn{3}{c|}{29.295} &
  \multicolumn{3}{c|}{29.36} &
  \multicolumn{3}{c|}{\bf 28.709} \\ \cline{2-13} 
\end{tabular}
}
\caption{KID scores for sampled axial, coronal and sagittal slices from the generated isotropic volume in the brain dataset. The table includes averaged KID scores for each plane individually and the overall average KID score across all planes for each case.}
\label{table:res_oasis}
\end{table}

\subsubsection*{Distribution-Based Metrics:} Table~\ref{table:res_abdomen} and ~\ref{table:res_oasis} present the metric scores for each of the methods applied to the abdomen and brain datasets, respectively. In the abdomen dataset, SIMPLE scores lower than interpolation but higher than SMORE4 and averaged ATMEs in the coronal plane across all metrics. In the axial plane, SIMPLE achieves the lowest scores. SMORE4 and averaged ATME perform well in the coronal plane but struggle in the axial, resulting in a higher score ratio between the planes compared to the more balanced ratio for SIMPLE. For the brain dataset, SIMPLE produces slightly lower KID scores than SMORE4 and averaged ATMEs in both 2D and 3D evaluations, and significantly lower scores compared to interpolation. The most notable differences across all methods are observed in the coronal and axial planes.

\subsection{Ablation Study}
\label{sec:ablation_study}
Two ablation studies were conducted on the abdomen dataset to assess the individual contributions of each discriminator. Table~\ref{table:ablation_study} reports KID and IS scores for these models in comparison to SIMPLE. Model (a) achieved the lowest KID score in the coronal plane, while model (b) performed similarly to SIMPLE. IS scores across all models were comparable, with model (b) closest to SIMPLE. These results indicate that the axial discriminator contributes more significantly, likely due to the coronal plane’s higher resolution in the anisotropic data. The combined use of both discriminators in SIMPLE achieved the lowest average KID, emphasizing the importance of simultaneous optimization.

\begin{table}[t!]
\small
\resizebox{\columnwidth}{!}{
\begin{tabular}{ccccccc}
 &
   &
   &
  \multicolumn{4}{c}{\textbf{Mertic}} \\ \cline{4-7} 
 &
   &
  \multicolumn{1}{c|}{} &
  \multicolumn{2}{c|}{KID} &
  \multicolumn{2}{c|}{IS} \\ \cline{2-7} 
\multicolumn{1}{c|}{\textbf{Method}} &
  \multicolumn{1}{c|}{\begin{tabular}[c]{@{}c@{}}Coronal \\ Disc\end{tabular}} &
  \multicolumn{1}{c|}{\begin{tabular}[c]{@{}c@{}}Axial\\ Disc\end{tabular}} &
  \multicolumn{1}{c|}{\textit{Coronal}} &
  \multicolumn{1}{c|}{\textit{Axial}} &
  \multicolumn{1}{c|}{\textit{Coronal}} &
  \multicolumn{1}{c|}{\textit{Axial}} \\ \cline{2-7} 
\multicolumn{1}{c|}{\multirow{2}{*}{(a)}} &
  \multicolumn{1}{c|}{\multirow{2}{*}{\checkmark}} &
  \multicolumn{1}{c|}{\multirow{2}{*}{-}} &
  \multicolumn{1}{c|}{0.268} &
  \multicolumn{1}{c|}{39.597} &
  \multicolumn{1}{c|}{0.0027} &
  \multicolumn{1}{c|}{0.0219} \\ \cline{4-7} 
\multicolumn{1}{c|}{} &
  \multicolumn{1}{c|}{} &
  \multicolumn{1}{c|}{} &
  \multicolumn{2}{c|}{19.932} &
  \multicolumn{2}{c|}{0.0123} \\ \cline{2-7} 
\multicolumn{1}{c|}{\multirow{2}{*}{(b)}} &
  \multicolumn{1}{c|}{\multirow{2}{*}{-}} &
  \multicolumn{1}{c|}{\multirow{2}{*}{\checkmark}} &
  \multicolumn{1}{c|}{4.218} &
  \multicolumn{1}{c|}{32.451} &
  \multicolumn{1}{c|}{0.0031} &
  \multicolumn{1}{c|}{0.0047} \\ \cline{4-7} 
\multicolumn{1}{c|}{} &
  \multicolumn{1}{c|}{} &
  \multicolumn{1}{c|}{} &
  \multicolumn{2}{c|}{18.334} &
  \multicolumn{2}{c|}{0.0039} \\ \cline{2-7} 
\multicolumn{1}{c|}{\multirow{2}{*}{SIMPLE}} &
  \multicolumn{1}{c|}{\multirow{2}{*}{\checkmark}} &
  \multicolumn{1}{c|}{\multirow{2}{*}{\checkmark}} &
  \multicolumn{1}{c|}{2.383} &
  \multicolumn{1}{c|}{32.487} &
  \multicolumn{1}{c|}{0.0027} &
  \multicolumn{1}{c|}{0.0058} \\ \cline{4-7} 
\multicolumn{1}{c|}{} &
  \multicolumn{1}{c|}{} &
  \multicolumn{1}{c|}{} &
  \multicolumn{2}{c|}{{\bf 17.435}} &
  \multicolumn{2}{c|}{{\bf 0.0042}} \\ \cline{2-7} 
\end{tabular}
}
\caption{Ablation study results. The full SIMPLE model results compared to partial models based on a single coronal (a) or axial (b) discriminator.}
\label{table:ablation_study}
\end{table}
\section{Conclusion}
\label{sec:conclusion}
We introduce SIMPLE, a self-supervised approach for isotropic MRI restoration from anisotropic data using multi-plane clinical images. SIMPLE outperforms state-of-the-art methods in quantitative metrics and radiologist evaluations, enhancing volumetric analysis and 3D reconstructions with potential clinical benefits. Future work will focus on optimizations and applications to other imaging modalities.



{
    \small
    \bibliographystyle{ieeenat_fullname}
    \bibliography{main}
}

\clearpage
\setcounter{page}{1}
\maketitlesupplementary

\begin{figure}[H]
  \centering
  \begin{minipage}{0.85\textwidth}  
    \centering
  \rotatebox{270}{ 
    \parbox{0.85\textheight}{ 
      \centering 
      \begin{tikzpicture}
          \node (img) {\includegraphics[height=0.29\textheight, width=1.04\linewidth]{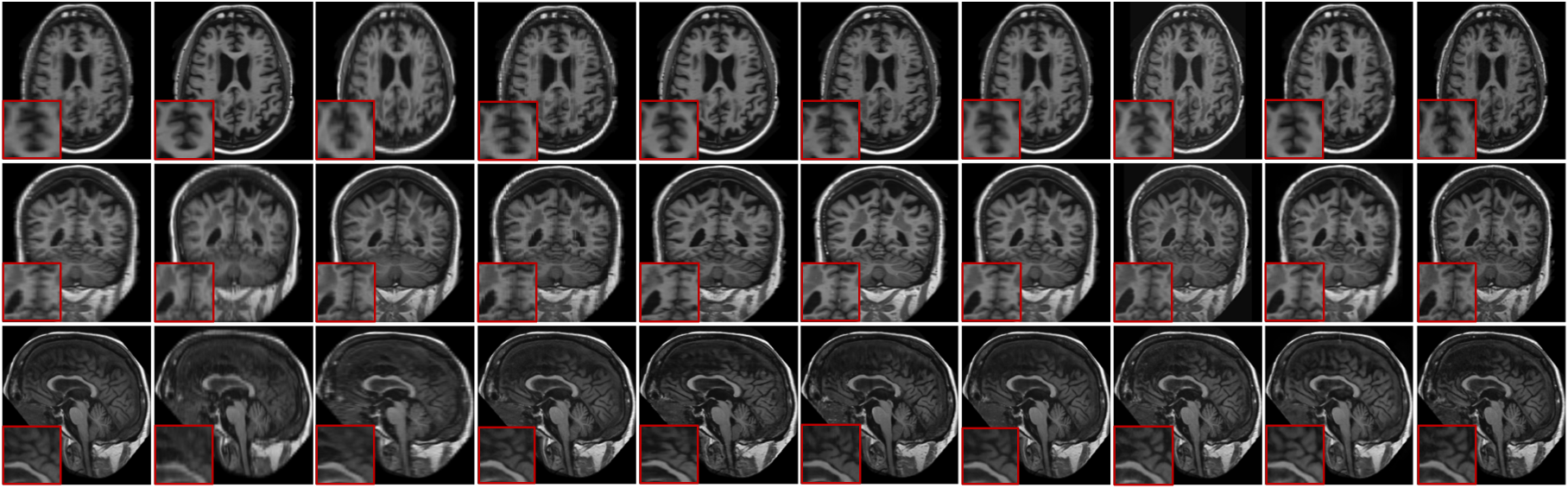}};
          
          \node[above] at ([xshift=-256,yshift=0pt]img.north) {\footnotesize Sagittal Volume};
          \node[above] at ([xshift=-199,yshift=0pt]img.north) {\footnotesize Axial Volume};
          \node[above] at ([xshift=-141,yshift=0pt]img.north) {\footnotesize Coronal Volume};
          \node[above] at ([xshift=-82,yshift=0pt]img.north) {\footnotesize Sagittal ATME};
          \node[above] at ([xshift=-24,yshift=0pt]img.north) {\footnotesize Axial ATME};
          \node[above] at ([xshift=34,yshift=0pt]img.north) {\footnotesize Coronal ATME};
          \node[above] at ([xshift=91,yshift=0pt]img.north) {\footnotesize Avg. ATME};
          \node[above] at ([xshift=146,yshift=0pt]img.north) {\footnotesize SMORE4};
          \node[above] at ([xshift=200,yshift=0pt]img.north) {\footnotesize SIMPLE};
          \node[above] at ([xshift=254,yshift=0pt]img.north) {\footnotesize GT};

          \rotatebox{90}{
          \node[left] at ([xshift=72pt,yshift=192pt]img.north) {\footnotesize \shortstack{Axial}};
          \node[left] at ([xshift=16pt,yshift=192pt]img.north) {\footnotesize \shortstack{Coronal}};
          \node[left] at ([xshift=-47pt,yshift=192pt]img.north) {\footnotesize \shortstack{Sagittal}};
          }
      \end{tikzpicture}

      \vspace{1em} 
      \parbox{1.36\textwidth}{\caption{Multi-plane views of slices from an isotropic MRI volume, derived from a brain dataset, generated by seven competing methods and compared with the ground-truth (GT) volume. From left to right: linear interpolation on anisotropic sagittal volume, linear interpolation on anisotropic axial volume, linear interpolation on anisotropic coronal volume, ATME on anisotropic sagittal volume, ATME on anisotropic axial volume, ATME on anisotropic coronal volume, averaged ATME of coronal, axial and sagittal planes on anisotropic sagittal volume, SMORE4 on anisotropic sagittal volume, and SIMPLE on anisotropic sagittal volume. Average ATME, SMORE4, and SIMPLE effectively enhance image quality and resolution across all planes, whereas linear interpolation and ATME demonstrate biases toward their original anisotropic acquisition plane. SIMPLE shows slightly better performance than Average ATME and SMORE4, achieving sharper images with less noise, fewer shadows, and enhanced contrast.}}
      \vspace{10em}
      \label{fig:multi_view_oasis}
    }
  }
  \end{minipage}
\end{figure}

\end{document}